\documentclass[journal, twocolumn]{IEEEtran}
\usepackage{amsmath,amsfonts}
\usepackage{algorithmic}
\usepackage{algorithm}
\usepackage{array}
\usepackage[caption=false,font=normalsize,labelfont=sf,textfont=sf]{subfig}
\usepackage{textcomp}
\usepackage{stfloats}
\usepackage{url}
\usepackage{verbatim}
\usepackage{graphicx}
\usepackage{cite}
\usepackage{amsthm}
\usepackage{hyperref}
\usepackage{algorithm}
\usepackage{amssymb}
\usepackage{bbm}
\usepackage{makecell}
\usepackage{multirow}
\usepackage{pgf}
\usepackage{xcolor, colortbl}

\newtheorem{theorem}{Theorem}
\newtheorem{corollary}[theorem]{Corollary}
\newtheorem{lemma}[theorem]{Lemma}
\newtheorem{proposition}[theorem]{Proposition}
\newtheorem{remark}{Remark}
\newtheorem{example}{Example}
\newtheorem{definition}{Definition}

\DeclareMathOperator*{\argmin}{arg\,min}

\begin{document}

\title{\huge Achieving the Exactly Optimal Privacy-Utility Trade-Off \\with Low Communication Cost via Shared Randomness}

\author{Seung-Hyun~Nam,~\IEEEmembership{Graduate Student Member,~IEEE,}
Hyun-Young~Park,~\IEEEmembership{Graduate Student Member,~IEEE,}
and Si-Hyeon~Lee,~\IEEEmembership{Senior Member,~IEEE}
\thanks{S.-H. Nam, H.-Y. Park, and S.-H. Lee are with the School of Electrical Engineering, Korea Advanced Institute of Science and Technology (KAIST), Daejeon 34141, South Korea  (e-mail: shnam@kaist.ac.kr; phy811@kaist.ac.kr; sihyeon@kaist.ac.kr).
S.-H. Nam and H.-Y. Park contributed equally to this work. (Corresponding Author: Si-Hyeon Lee)}
}

\maketitle

\begin{abstract}
    We consider a discrete distribution estimation problem under a local differential privacy (LDP) constraint in the presence of shared randomness.
    By exploiting the shared randomness, we suggest a new method for constructing LDP schemes which achieve the {exactly} optimal privacy-utility trade-off (PUT) with the communication cost of less than or equal to the input data size for any privacy regime.
    The main idea is to decompose a block design scheme by Park et al. (2023), based on the combinatorial concept called resolution.
    The LDP scheme decomposed from a block design scheme is called a resolution of the block design scheme, and it achieves the same {PUT} as the original block design scheme while requiring a less communication cost. 

We provide two resolutions of an exactly PUT-optimal block design scheme, called the Baranyai's resolution and the cyclic shift resolution, both requiring the communication cost of less than or equal to the input data size.
In particular, we show that the Baranyai's resolution achieves the minimum communication cost among all {the PUT-optimal resolutions of block design schemes.}
One drawback of the Baranyai's resolution is that it can be obtained through a recursive algorithm in general.
In contrast, the cyclic shift resolution has an explicit structure, but its communication cost can be larger than that of Baranyai's resolution.
To complement this, we also suggest resolutions of other block design schemes achieving the optimal PUT for some privacy budgets, which require the minimum communication cost as the Baranyai's resolution and have explicit structures as the cyclic shift resolution.
    
\end{abstract}

\section{Introduction}
In statistical inference problems, local differential privacy (LDP) is one of the most representative  privacy requirements for preserving privacy during the data collection process \cite{kasiviswanathan11_learn_private,duchi13_LDP,issa19_operational}.
The LDP requires that {the distribution of} the data sent to the server is not significantly different depending on the original data, making the server hard to infer the original data from the collected data.
To satisfy LDP, the original data is in general perturbed randomly through a conditional distribution, called a privacy mechanism. Due to such a perturbation, there is a fundamental trade-off between the privacy leakage and the inference error, called privacy-utility trade-off (PUT).
The PUTs have been studied for various inference tasks such as discrete distribution estimation \cite{duchi13_LDP, ye18_SS, acharya19_HR, barnes20_ldp_fisher, Feldman22_PGR}, mean estimation \cite{duchi13_LDP, wang19_piecewise, asi22_Privunit_opt, bhowmick18_privunit}, and federated learning \cite{bhowmick18_privunit, truex20_ldpfed,zhao20_ldpfed_iot,seif20_wireless_ldpfed}.
For the discrete distribution estimation task which we focus on in this paper, the subset selection (SS) scheme proposed in \cite{ye18_SS} is known to achieve the exactly optimal PUT \cite{ye19_SSOptim, park23_block}.
However, the SS requires a high communication cost, whose order is greater than the exponential of the input data size.


In practice, the communication cost of a privacy mechanism is an important factor together with the PUT. 
Many communication-efficient LDP schemes have been proposed for discrete distribution estimation with and without shared randomness between the clients and the server \cite{acharya19_HR,Feldman22_PGR,park23_block,chen20_trilemma,shah22_MMRC,bassily15local,wang17_LDPFE}.  
For the works without shared randomness, the Hadamard response \cite{acharya19_HR} and the projective geometry response \cite{Feldman22_PGR} were proposed based on the  Hadamard matrix and based on the finite projective geometry, respectively, while both the works did not guarantee that their schemes achieve the exactly-optimal PUT.
A more recent work \cite{park23_block} proposed a class of LDP schemes without shared randomness, called block design schemes. This class of block design schemes  not only unifies lots of previously suggested schemes, e.g., the randomized response \cite{stanley65_RR}, the Hadamard response \cite{acharya19_HR}, the projective geometry response \cite{Feldman22_PGR}, and the exactly PUT-optimal SS  \cite{ye18_SS}, through the lens of combinatorial block designs \cite{stinson04_combinatorial,ionin06_combSymDesign,colbourne07_handbook}, but also suggests a novel way of constructing exactly PUT-optimal schemes with lower communication costs.
Indeed, by considering this class of block design schemes, the exactly optimal PUT was shown to be achieved by using the minimum communication cost\footnote{In this paper,  we consider the minimum communication cost to be able to construct a consistent estimator, where we say an estimator is consistent if its estimation error converges to zero as the number of data increases.} given as the input data size \cite{acharya19_PUCT}, for a new regime of input data size and privacy budget not covered previously. However, due to the sparse existence of combinatorial block designs, for some pairs of input data size and LDP budget, an exactly PUT-optimal block design scheme with communication cost equal to the input data size may not exist.

There have been a line of research exploiting 
shared randomness to reduce the communication cost  \cite{bassily15local, wang17_LDPFE, chen20_trilemma, shah22_MMRC}. The shared randomness between each client and the server can be available when the server generates the randomness and sends it to the clients prior to data collection, and it is also equivalent to the assumption that the LDP {mechanism} of each client can be separately designed and informed to  the server. 
By exploiting shared randomness, the work \cite{bassily15local} suggested a method to convert any given LDP scheme without shared randomness into an LDP scheme with shared randomness which has communication cost of 1-bit without loss in the order of PUT, and the work \cite{wang17_LDPFE} reduced the communication cost of a variant of RAPPOR \cite{erlingsson14_rappor} based on shared random projection matrices.
The recursive Hadamard response (RHR) \cite{chen20_trilemma} was shown to achieve the order-optimal privacy-utility-communication trade-off, and the work \cite{shah22_MMRC} suggested a compressed version of SS which achieves a better trade-off than RHR.
We note that these prior works on LDP schemes with low communication costs exploiting shared randomness either show the optimality in the order-optimal sense or assess the superiority based on numerical simulations.



In this paper, we suggest a new method for constructing LDP schemes with shared randomness which achieve the exactly optimal PUT with the communication cost of less than or equal to the input data size, for all values of input data size and LDP budget.  Here, the exactly optimal PUT is guaranteed as the SS scheme \cite{ye19_SSOptim, park23_block}, not in the order-optimal sense.
The main idea is to decompose a PUT-optimal block design scheme by exploiting shared randomness, based on the combinatorial concept called resolution \cite{stinson04_combinatorial, ionin06_combSymDesign}.
We call the LDP scheme decomposed from a block design scheme via shared randomness a resolution of the block design scheme.
A resolution of a block design scheme  achieves the same  PUT as the original block design scheme, while requiring a less communication cost. 

In this work, we provide two resolutions of the PUT-optimal SS, which are from the Baranyai's theorem \cite{Baranyai74_Baranyai} and from the cyclic shift group action, both requiring the communication cost of less than or equal to the input data size.
In particular, we show that the Baranyai's resolution achieves the minimum communication cost among all the {PUT-optimal resolutions of block design schemes.}
One drawback of the Baranyai's resolution is that it can be obtained through a recursive algorithm in general \cite{Baranyai74_Baranyai,chee21_Baranyai_const4}.
In contrast, the cyclic shift resolution has an explicit structure, but its communication cost can be larger than that of Baranyai's resolution.
To complement this, we also suggest resolutions of other block design schemes achieving the optimal PUT for some cases of privacy budget, which require the minimum communication cost as the Baranyai's resolution and have explicit structures as the cyclic shift resolution.

The rest of the paper is organized as follows.
The problem of discrete distribution estimation under LDP constraint with shared randomness is formulated in Section~\ref{sec:form}, and 
some preliminary results on combinatorics and the class of block design schemes are reviewed in Section~\ref{sec:pre}. 
In Section~\ref{sec:res}, we present the proposed method of decomposing a block design scheme by exploiting shared randomness.
We present two decompositions of the PUT-optimal SS in Section~\ref{sec:res_SS}, and two decompositions of other block design schemes in Section \ref{sec:other}. Finally, Section~\ref{sec:conc} concludes the paper.

\section{Problem Formulation}\label{sec:form}
\subsection{Notations}
For integers $a<b$, $[a:b]$ denotes $\{c \in \mathbb{Z} : a \leq c \leq b\}$.
For a positive integer $v$, $\Delta_v \subset \mathbb{R}^v$ denotes a probability simplex,
\begin{equation}
    \Delta_v:=\left\{x \in \mathbb{R}^v : x_1,\cdots,x_v \geq 0,\; \sum\limits_{i=1}^v x_i = 1 \right\}.    
\end{equation}
We write $\log$ for the logarithm with base 2, and $\ln$ for the natural logarithm.
For a discrete probability distribution $P$, $H(P)$ denotes the entropy (in bits) of a random variable following the distribution $P$.

\subsection{System model}
\begin{figure}
    \centering
    \includegraphics[width=0.7\linewidth]{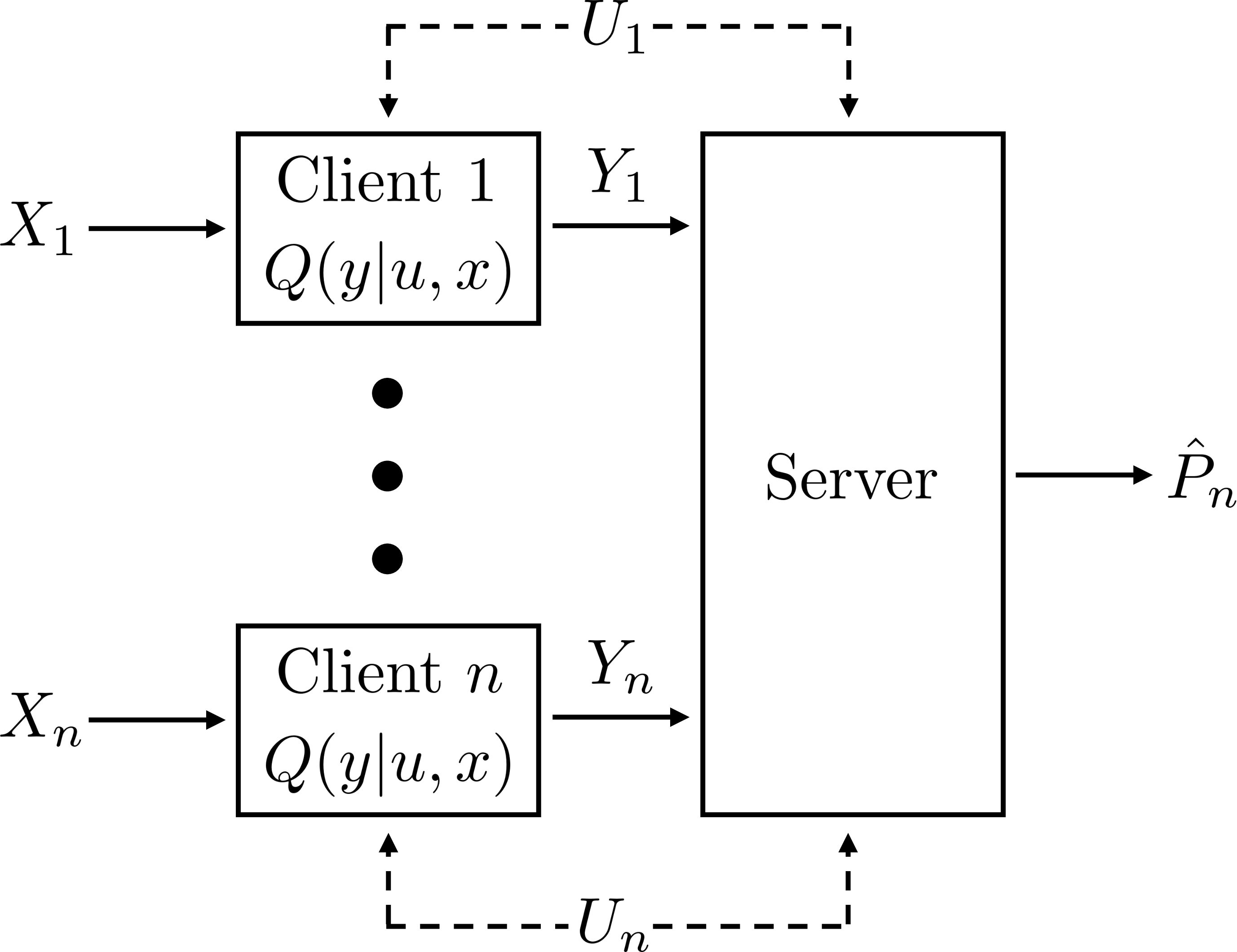}
    \caption{A discrete distribution estimation problem with shared randomness under LDP constraint. A privacy mechanism $Q$ should satisfy LDP constraint.}
    \label{fig:model}
\end{figure}
We consider the discrete distribution estimation problem depicted in Fig. \ref{fig:model}, where a server wants to estimate the underlying distribution of clients' data under LDP constraint when the server and each client have access to their own shared randomness.
There are $n$ clients and the $i$-th client for $i\in [1:n]$ has a data $X_i \in \mathcal{X} = [1:v]$ that is i.i.d. according to an unknown distribution $P_X \in \Delta_v$.
The shared randomness $U_i \in \mathcal{U}$ between the $i$-th client and the server is i.i.d. according to a pre-designed distribution $P_U$ on a finite set $\mathcal{U}$ and $U_i$'s are independent to $X_i$'s.\footnote{As discussed  in \cite[Sec. B]{chen20_trilemma}, one possible way to generate and share the randomness is that the server generates $U_1,\cdots,U_n$  i.i.d. according to $P_U$ and sends them to the clients in advance.
Also, the assumption of available shared randomness is equivalent to the assumption that the LDP scheme of each client can be separately designed and informed to the server.}
To mitigate privacy leakage, each client perturbs $X_i$ into $Y_i \in \mathcal{Y}$, where $\mathcal{Y}$ is assumed to be finite, through a conditional distribution $Q(y|u,x):\mathcal{U}\times\mathcal{X} \rightarrow \mathcal{Y}$ and the shared randomness $U_i$.
We call a pair $(P_U,Q)$  a\textbf{ privacy mechanism with shared randomness}.
Without loss of generality, we assume $P_U(u) > 0$ for all $u \in \mathcal{U}$, 
\begin{equation}
    \forall y,\; \exists u,x, \; Q(y|u,x)>0,
\end{equation}
and denote
\begin{equation}
    \mathcal{Y}_u = \{y \in \mathcal{Y}: Q(y|u,x) > 0 \text{ for some }x \in \mathcal{X}\},    
\end{equation}
and $\mathcal{Y}_i := \mathcal{Y}_{U_i}$.
We consider privacy mechanisms with shared randomness satisfying $\epsilon$-LDP constraint, defined as follows.
\begin{definition}[Local differential privacy]
    For $\epsilon>0$, a privacy mechanism with shared randomness $(P_U,Q)$ is said to satisfy $\epsilon$-LDP constraint  and is called an $\epsilon$-\textbf{LDP mechanism with shared randomness} if
    \begin{multline}
        \forall u \in \mathcal{U}, A \subset \mathcal{Y}_u,\;  x,x' \in \mathcal{X},
        \\ Q(A|u,x) \leq e^\epsilon Q(A|u,x').
    \end{multline} 
\end{definition}
The server collects privatized data $Y_1,\cdots,Y_n$ and estimates the underlying distribution $P_X$ through an estimator ${\hat{P}_n:(\mathcal{U} \times \mathcal{Y})^n \rightarrow \mathbb{R}^v}$, $\hat{P}_n = (\hat{P}_{n,1},\cdots,\hat{P}_{n,v})$.
A pair of an $\epsilon$-LDP mechanism with shared randomness and an estimator, $((P_U,Q),\hat{P}_n)$, is called an \textbf{$\epsilon$-LDP scheme with shared randomness}.
Because $P_X$ is unknown, we consider  the following worst-case mean squared error as the \textbf{estimation loss}:
\begin{equation}
    L_n(v,(P_U,Q),\hat{P}_n) := \sup\limits_{P_X \in \Delta_v} \mathbb{E}\left[ ||P_X-\hat{P}_n||_2^2 \right].
\end{equation}
Thus, the optimal privacy-utility trade-off (PUT) is defined as
\begin{equation}
    M_n(v,\epsilon):=\inf\limits_{((P_U,Q),\hat{P}_n)} L_n(v,(P_U,Q),\hat{P}_n),
\end{equation}
where the infimum is over the set of all $\epsilon$-LDP schemes with shared randomness.
We also say the estimation loss of a given $\epsilon$-LDP scheme with shared randomness as the PUT achieved by the scheme.

The following remark states that the use of shared randomness does not improve the optimal PUT.

\begin{remark}\label{rmk:shared_rand}
    A system without shared randomness can be interpreted as a special case of our system with $|\mathcal{U}|=1$.
    In this case, we simply denote a privacy mechanism as $\tilde{Q}:\mathcal{X} \rightarrow \mathcal{Z}$, an estimator as $\tilde{P}_n:\mathcal{Z}^n \rightarrow \mathbb{R}^v$, and the corresponding estimation loss as $L_n(v,\tilde{Q},\tilde{P}_n)$.
     For simplicity, we call a mechanism (or scheme) without shared randomness as a mechanism (or scheme) if there is no confusion.
    We note that the optimal PUT over $\epsilon$-LDP schemes with shared randomness is equal to the optimal PUT over $\epsilon$-LDP schemes  without shared randomness.
    To see this, for a given $\epsilon$-LDP scheme with shared randomness $((P_U,Q), \hat{P}_n)$, define a conditional distribution $\tilde{Q}(u,y|x) := Q(y|u,x)P_U(u)$.
    Then, it is easy to check that $\tilde{Q}:\mathcal{X} \rightarrow \mathcal{U} \times \mathcal{Y}$ is an $\epsilon$-LDP mechanism and by letting $\mathcal{Z}=\mathcal{U}\times \mathcal{Y}$ and $\tilde{P}_n = \hat{P}_n$, we have
    \begin{equation}
        L_n(v,(P_U,Q),\hat{P}_n) = L_n(v,\tilde{Q},\tilde{P}_n).
    \end{equation}
    Hence, we can conclude that the PUT achieved by an arbitrary $\epsilon$-LDP scheme with shared randomness can be achieved by an $\epsilon$-LDP scheme without shared randomness.    
\end{remark}

In this work, we are interested in achieving the optimal PUT with reduced communication cost from the clients to the server by utilizing shared randomness.
Since both the $i$-th client and the server know $U_i$ for all $i$, 
the server knows that $Y_i$ must be in $\mathcal{Y}_i$.
However, the server does not know how $Y_i$ is distributed over   $\mathcal{Y}_{i}$ because $P_X$ is unknown.
Hence, we assume that each client uses the uncoded transmission for $\mathcal{Y}_{i}$, which requires $\log |\mathcal{Y}_{i}|$-bits.
Thus, we define the (average) \textbf{communication cost} of an $\epsilon$-LDP mechanism with shared randomness $(P_U,Q)$ as follows: 
\begin{equation}\label{eq:comm_cost}
    C(P_U,Q):=\sum_{u \in \mathcal{U}}P_U(u) \log |\mathcal{Y}_{u}| ~\mbox{bits}.
\end{equation}


Our goal is to find communication-efficient PUT-optimal LDP schemes by utilizing shared randomness.

\section{Preliminaries} \label{sec:pre}
In this section, we present some preliminary results in combinatorics and review the class of LDP schemes proposed in \cite{park23_block}, on which our development is based.

\subsection{Block design}
In this subsection, we briefly review combinatorial block designs and their properties as preliminaries.
The details can be found in \cite{stinson04_combinatorial,ionin06_combSymDesign,colbourne07_handbook}.

\begin{definition}
An \textbf{incidence structure} is a triplet $(\mathcal{X},\mathcal{Z},\mathcal{I})$, where $\mathcal{X}$, $\mathcal{Z}$ are sets, and $\mathcal{I} \subset \mathcal{X} \times \mathcal{Z}$.
Furthermore, we define 
\begin{IEEEeqnarray}{rCl}
    \mathcal{I}_x &:=& \{z \in \mathcal{Z} : (x,z) \in \mathcal{I}\},\\
    \mathcal{I}^z &:=& \{x \in \mathcal{X} : (x,z) \in \mathcal{I}\}. 
\end{IEEEeqnarray}
\end{definition}
\begin{definition}
    Two incidence structures $(\mathcal{X},\mathcal{Z},\mathcal{I})$ and $(\mathcal{X}',\mathcal{Z}',\mathcal{I}')$ are called isomorphic if there exist bijections $f:\mathcal{X} \rightarrow \mathcal{X}'$ and $g:\mathcal{Z} \rightarrow \mathcal{Z}'$ such that
    \begin{equation}
        (x,z) \in \mathcal{I} \Leftrightarrow (f(x),g(z)) \in \mathcal{I}'.
    \end{equation}
\end{definition}

In this paper, we focus on incidence structures with $|\mathcal{X}|,|\mathcal{Z}| < \infty$.
Also, the results in this paper only rely on some properties of incidence structures which are preserved under isomorphism.
\begin{definition}
Let $(\mathcal{X},\mathcal{Z},\mathcal{I})$ be an incidence structure. We say this incidence structure is
\begin{itemize}
    \item $r$-regular, if $ \forall x \in \mathcal{X},\; |\mathcal{I}_x|=r$.
    \item $k$-uniform, if $\forall z \in \mathcal{Z},\; |\mathcal{I}^z|=k$.
    \item $\lambda$-{pairwise balanced}, if $|\mathcal{I}_{x} \cap \mathcal{I}_{x'}|=\lambda$, for all ${x,x' \in \mathcal{X}}$, $x \neq x'$.
\end{itemize}    
\end{definition}

\begin{definition}
Let $v,b,r,k$, and $\lambda$ be integers such that ${v>k>0}$, $b>r>\lambda \geq 0$. A \textbf{block design} with parameters $(v,b,r,k,\lambda)$, or shortly $(v,b,r,k,\lambda)$-block design, is an incidence structure $(\mathcal{X},\mathcal{Z},\mathcal{I})$ such that $|\mathcal{X}|=v$, $|\mathcal{Z}|=b$, $r$-regular, $k$-uniform, and $\lambda$-pairwise balanced.
\end{definition}
\begin{lemma}\cite[Prop. 2.3.7]{ionin06_combSymDesign}\label{lem:blockDesignFundEq}
For any $(v,b,r,k,\lambda)$-block design, the parameters should satisfy
\begin{equation}
    vr=bk, \quad \lambda(v-1)=r(k-1).
\end{equation}
\end{lemma}
\begin{remark}\label{rmk:CBD}
    In this paper, we represent the complete block design with parameters $(v,k)$ as
    the incidence structure $(\mathcal{X},\mathcal{Z},\mathcal{I})$, where $\mathcal{X}=[1:v]$, $\mathcal{Z}= \{z \in \{0,1\}^v: \sum_i z_i = k\}$, and $\mathcal{I} = \{(x,z) \in \mathcal{X}\times \mathcal{Z}: z_x = 1\}$ where $z_x$ denotes the $x$-th element of vector $z$, among many other possible isomorphic representations.
    It is easy to check that a complete block design is a block design.
\end{remark}
The way that we construct communication-efficient PUT-optimal LDP mechanisms by exploiting shared randomness is related to a resolution of  an incidence structure defined as follows.
\begin{definition}
Let $(\mathcal{X}, \mathcal{Z}, \mathcal{I})$ be an incidence structure. 
\begin{itemize}
    \item A non-empty subset $\mathcal{C} \subset \mathcal{Z}$ is called a \textbf{resolution class} if there exists a positive integer $\alpha(\mathcal{C})$ such that
    \begin{equation}\label{eq:res_class}
        \forall x \in \mathcal{X},\quad  |\mathcal{I}_x \cap \mathcal{C}| = \alpha(\mathcal{C}).
    \end{equation}
    \item A \textbf{resolution} of $(\mathcal{X}, \mathcal{Z}, \mathcal{I})$ is a partition $\mathcal{R}$ of $\mathcal{Z}$ which consists of resolution classes.
    If $\mathcal{R}$ satisfies $\alpha(\mathcal{C})=\alpha$, $\forall \mathcal{C} \in \mathcal{R}$, then we say such resolution is an $\alpha$-resolution.
    \item An incidence structure is said to be $\alpha$-resolvable if there is an $\alpha$-resolution of it.
    If an incidence structure is $1$-resolvable, then we simply say that it is \textbf{resolvable}.
\end{itemize}
\end{definition}
\begin{lemma}\cite[Ch. 5]{ionin06_combSymDesign}\label{lem:res_lem}
\begin{enumerate}
    \item If an incidence structure has a resolution, then it  is regular.
    \item For any resolution class $\mathcal{C}$ of a $(v,b,r,k,\lambda)$-block design, we have $k|\mathcal{C}|=v\alpha(\mathcal{C})$.
    \item For any resolution $\mathcal{R}$ of $(v,b,r,k,\lambda)$-block design, we have $\sum_{\mathcal{C} \in \mathcal{R}} |\mathcal{C}|=b$ and $\sum_{\mathcal{C} \in \mathcal{R}} \alpha(\mathcal{C})=r$.
\end{enumerate}

\end{lemma}

\subsection{Block design scheme}
In this subsection, we introduce the block design scheme proposed in \cite{park23_block}  with some additional comments.

\begin{definition}\label{def:blockDesignMech}
For given $\mathcal{X}=\{1,2,\cdots,v\}$ and $\epsilon>0$, an $\epsilon$-LDP mechanism $\tilde{Q}$ is called a \textbf{block design mechanism} with parameters $(v,b,r,k,\lambda,\epsilon)$ if there exists a $(v,b,r,k,\lambda)$-block design $(\mathcal{X},\mathcal{Z},\mathcal{I})$ such that
    \begin{equation}\label{eq:mech}
        \tilde{Q}(z|x)=\begin{cases} \frac{e^\epsilon}{(e^\epsilon-1)r+b} & \text{ if }(x,z) \in \mathcal{I} \\ \frac{1}{(e^\epsilon-1)r+b} & \text{ otherwise} \end{cases}.
    \end{equation}
\end{definition}

The following theorem states that there is a natural unbiased estimator for a block design mechanism, and the estimation loss of a pair of block design mechanism and the corresponding unbiased estimator can be simply represented as a function of $(n,v,k,\epsilon)$.
\begin{theorem}\cite[Thm. 3.2]{park23_block}\label{thm:BDML2risk}
For given $n\in \mathbb{N}$ and a $(v,b,r,k,\lambda,\epsilon)$-block design mechanism $\tilde{Q}$, there is an unbiased estimator given as
\begin{multline}\label{eq:Canonical est}
    \tilde{P}_{n,x}(Z_1,\cdots,Z_n) = \frac{(v-1)(ke^\epsilon + v-k)}{nk(v-k)(e^\epsilon-1)}N_x(Z_1,\cdots,Z_n)
     \\  -\frac{(k-1)e^\epsilon+v-k}{(v-k)(e^\epsilon-1)} ,
\end{multline}
where
\begin{align}\label{eq:Nx}
N_x(Z_1,\cdots,Z_n) = \sum_{i=1}^{n}\mathbbm{1}(Z_i \in \mathcal{I}_x).
\end{align}
Moreover, the estimation loss is given by
\begin{equation}
    L_n(v,\tilde{Q},\tilde{P}_n)=\frac{(v-1)^2(ke^\epsilon+v-k)^2}{nvk(v-k)(e^\epsilon-1)^2}. \label{eq:BDWorstCaseRisk}
\end{equation}
\end{theorem}

We call a pair of block design mechanism and the corresponding  unbiased estimator \eqref{eq:Canonical est} as a \textbf{block design scheme}.
\begin{corollary}
    If $\tilde{Q}$ and $\tilde{Q}'$ are block design mechanisms which correspond to block designs having the same parameters $(v,k,\epsilon)$, then the block design schemes based on $\tilde{Q}$ and $\tilde{Q}'$ achieve the same estimation loss.
\end{corollary}
Because isomorphic block designs have the same parameters, it is easy to see that the block design schemes based on isomorphic block designs achieve the same estimation loss.
\begin{remark}
    The SS with parameters $(v,k,\epsilon)$ is the $\epsilon$-LDP block design scheme based on $\left(v,k \right)$-complete block design \cite{ye18_SS ,park23_block}.
    The optimal SS \cite{ye19_SSOptim} which achieves the optimal PUT has the parameters $(v,k^*,\epsilon)$, where 
    \begin{equation}
        k^* \in \argmin_k \frac{(ke^\epsilon+v-k)^2}{k(v-k)}.
    \end{equation}
    It is easy to check that $k^* \in \left\{ \lfloor \frac{v}{e^\epsilon+1} \rfloor, \lceil \frac{v}{e^\epsilon+1} \rceil \right\}$.
    Further, the following proposition presents a stronger optimality condition.
    \begin{proposition}\cite[Prop. 3.6]{park23_block}\label{prop:optimal_eqcond} 
        A block design scheme with parameters $(v,k,\epsilon)$ achieves the optimal PUT if and only if 
        \begin{equation}
            E(k,k+1;v) \leq \epsilon \leq E(k-1,k;v),\label{eq:opt_range}
        \end{equation}
        where
        \begin{equation}
            E(k_1,k_2;v):= \frac{1}{2} \ln \frac{(v-k_1)(v-k_2)}{k_1 k_2}, 
        \end{equation}
        and $E(0,1;v):=\infty$.
        Further, let $K_{opt}(v,\epsilon)$ be the set of $k$'s satisfying \eqref{eq:opt_range}.
        If there exists $t \in \mathbb{Z}$ such that $\epsilon = {E(t,t+1;v)}$, then $K_{opt}(v,\epsilon)=\{t,t+1\}$. Otherwise, $K_{opt}(v,\epsilon)$ is a singleton.
    \end{proposition}
    Proposition \ref{prop:optimal_eqcond} says that the optimal uniformity parameter $k^*$ for SS that achieves the optimal PUT {might not be unique}.
    In this paper, we choose the following uniformity parameter as the optimal one (among possibly multiple optimal uniformity parameters):
    \begin{equation}
        k^* := \argmin\limits_{k \in K_{opt}(v,\epsilon)} \frac{v}{\mathrm{gcd}(v,k)}.
    \end{equation}
\end{remark}

As mentioned in \cite[Sec. 1.3]{stinson04_combinatorial}, there is a natural one-to-one correspondence (up to isomorphism) between incidence structures and $\{0,1\}$-valued matrices, called incidence matrices.
\begin{definition}
    A \{0,1\}-valued $v \times b$ matrix $M$ is an incidence matrix of an incidence structure $(\mathcal{X},\mathcal{Z},\mathcal{I})$ if the incidence structure $([1:v],[1:b],\tilde{I})$ with
    \begin{equation}
        \tilde{I}:= \{(i,j): M_{ij}=1,\;i \in [1:v],\; j \in [1:b]\},
    \end{equation}
    is isomorphic to $(\mathcal{X},\mathcal{Z},\mathcal{I})$.
\end{definition}
Treating an incidence structure as an incidence matrix is useful to get intuitions about block design schemes.
For a given incidence matrix $M$ of a block design, the stochastic matrix of the corresponding block design mechanism can be constructed by simply substituting $0\mapsto 1$, $1 \mapsto e^\epsilon$ in $M$ and taking normalization so that the each row sums to one.
Also, note that a resolution of a block design can be treated as a partition of columns of the corresponding incidence matrix such that the row sums of the submatrix formed by each set of columns is a constant.

\begin{example} \label{ex:42ss}
    A matrix $M$ is an incidence matrix of the $(4,2)$-complete block design if
    \begin{equation}\label{eq:42cbd}
        M = \begin{pmatrix}
        1 & 1 & 1 & 0 & 0 & 0 \\
        1 & 0 & 0 & 1 & 1 & 0 \\
        0 & 1 & 0 & 1 & 0 & 1 \\
        0 & 0 & 1 & 0 & 1 & 1 
        \end{pmatrix}.
    \end{equation}
    By replacing $0\mapsto 1$, $1 \mapsto e^\epsilon$ in $M$, we get
    \begin{equation}
        \tilde{M} = \begin{pmatrix}
        e^\epsilon & e^\epsilon & e^\epsilon & 1 & 1 & 1 \\
        e^\epsilon & 1 & 1 & e^\epsilon & e^\epsilon & 1 \\
        1 & e^\epsilon & 1 & e^\epsilon & 1 & e^\epsilon \\
        1 & 1 & e^\epsilon & 1 & e^\epsilon & e^\epsilon 
        \end{pmatrix}.
    \end{equation}
    Because the row sums of an incidence matrix of a block design are the same by the regularity, we can get a  stochastic matrix $\tilde{Q}$ by normalizing $\tilde{M}$,
    \begin{equation}\label{eq:42ss}
        \tilde{Q} = \frac{1}{3(e^\epsilon+1)} \begin{pmatrix}
        e^\epsilon & e^\epsilon & e^\epsilon & 1 & 1 & 1 \\
        e^\epsilon & 1 & 1 & e^\epsilon & e^\epsilon & 1 \\
        1 & e^\epsilon & 1 & e^\epsilon & 1 & e^\epsilon \\
        1 & 1 & e^\epsilon & 1 & e^\epsilon & e^\epsilon 
        \end{pmatrix}.
    \end{equation}
Note that the above row stochastic matrix $\tilde{Q}$ corresponds to the  $(4,2,\epsilon)$-SS mechanism.
\end{example}


Based on the fact that the estimation loss of a block design scheme only depends on $(v,k,\epsilon)$, it can be  shown that any block design scheme with the parameters $(v,k^*,\epsilon)$ achieves the optimal PUT among all the $\epsilon$-LDP schemes asymptotically, and the optimal PUT among all the unbiased $\epsilon$-LDP schemes non-asymptotically \cite[Thm. 3.1]{ye19_SSOptim}, \cite[Thm. 3.5, Thm. 3.7]{park23_block}.
In terms of the communication cost, the optimal SS requires the communication cost of $\Omega(v)$-bits, at least exponential in the input data size of $\log v$-bits.
To reduce the communication cost, the previous work \cite{park23_block} found some symmetric block design schemes\footnote{A $(v,b,r,k,\lambda)$-block design is called symmetric if $v=b$.} with parameters $(v,k^*,\epsilon)$ which achieve the optimal PUT with communication cost of $\log v$-bits.
We note that  the communication cost of $\log v$-bits is the minimum communication cost for a consistent estimator to exist in the absence of shared randomness \cite{acharya19_PUCT}.
However, depending on the values of $(v,\epsilon)$, such symmetric block design schemes may not exist.

\section{Resolution of a Block Design Scheme} \label{sec:res}
In this section, we propose a way to reduce the communication cost of a block design scheme by using shared randomness.
First, we define a decomposition of an LDP mechanism by using shared randomness.
The basic idea is to decompose an output of a mechanism without shared randomness into two parts: a part that can be constructed by shared randomness and the other part which the mechanism should transmit.

\begin{definition}\label{def:decomp}
An $\epsilon$-LDP mechanism with shared randomness $(P_U, {Q})$, $Q:\mathcal{U} \times \mathcal{X} \rightarrow \mathcal{Y}$, is a \textbf{decomposition} of an $\epsilon$-LDP mechanism without shared randomness $\tilde{Q}:\mathcal{X} \rightarrow \mathcal{Z}$ if there exists an injection $f_u:\mathcal{Y}_u \rightarrow \mathcal{Z}$ for each $u \in \mathcal{U}$ such that
\begin{multline}\label{eq:decomposition}
    \forall x \in \mathcal{X},\; z \in \mathcal{Z},
\\ \tilde{Q}(z|x) = \sum_{u \in \mathcal{U}} \sum\limits_{\substack{y \in \mathcal{Y}_u : \\ f_u(y) = z}} Q(y|u,x)P_U(u).
\end{multline}
\end{definition}


From the definition, the marginal distribution of output $Z$ from an $\epsilon$-LDP mechanism $\tilde{Q}$ and that of $f_U(Y)$ from its decomposition $(P_U,Q)$ are equal.
Thus, any estimator $\tilde{P}_n$ for $\tilde{Q}$ naturally induces an estimator $\hat{P}_n$ for $(P_U, Q)$ which preserves the estimation loss.
We state this fact in the following lemma whose proof is omitted.
\begin{lemma}\label{lem:est_dec}
Let $(P_U, Q)$ be a decomposition of an $\epsilon$-LDP mechanism without shared randomness $\tilde{Q}$.
For any estimator $\tilde{P}_n$ for $\tilde{Q}$, there is an estimator $\hat{P}_n$ for $(P_U, Q)$ given by
\begin{multline}
    \hat{P}_n((u_1,y_1),\cdots,(u_n,y_n))
    \\ = \tilde{P}_n(f_{u_1}(y_1),f_{u_2}(y_2),\cdots,f_{u_n}(y_n))
\end{multline}
for some injection $f_u:\mathcal{Y}_u \rightarrow \mathcal{Z}$ for each $u \in \mathcal{U}$, so that
\begin{align}
    L_n(v,(P_U,Q),\hat{P}_n) = L_n(v,\tilde{Q},\tilde{P}_n).
\end{align}
\end{lemma}

Note that the communication cost of $\tilde{Q}:\mathcal{X} \rightarrow \mathcal{Z}$ is $\log |\mathcal{Z}|$.
If $(P_U,Q)$ is a decomposition of $\tilde{Q}$, then the communication cost of $(P_U,Q)$ is less than or equal to that of $\tilde{Q}$ because
\begin{equation}
    C(P_U,Q) = \sum\limits_{u \in \mathcal{U}} P_U(u) \log |\mathcal{Y}_u| \leq \log |\mathcal{Z}|,
\end{equation}
where the inequality follows from the existence of an injection $f_u:\mathcal{Y}_u \rightarrow \mathcal{Z}$ for each $u \in \mathcal{U}$.
If an LDP mechanism is a block design mechanism whose underlying block design has a resolution, then such a block design mechanism can be decomposed by exploiting the  resolution.
We first illustrate through a simple example how a decomposition of a block design scheme can be constructed based on a resolution of the underlying block design.
\begin{example}\label{ex:res_block}
    Let $M$ be the matrix in \eqref{eq:42cbd}, which is an incidence matrix of the $(4,2)$-complete block design $(\mathcal{X},\mathcal{Z},\mathcal{I})$.
    Note that the set of columns of $M$ is equal to $\mathcal{Z}$.
    Let $\mathcal{U}=[1:3]$ and $\mathcal{Z}_u = \{M^u, M^{6-u}\}$, where $M^y$ denotes the $y$-th column of $M$.
    Then, it can be easily checked that $\{\mathcal{Z}_u\}_{u \in \mathcal{U}}$ is a resolution of $(\mathcal{X},\mathcal{Z},\mathcal{I})$.
    Now, let us construct $(Q,P_U)$.
    For the matrix $N_u = (M^u,M^{6-u})$, note that the row sums are equal from the definition of a resolution.  Thus, we can get a row stochastic matrix $Q_u$ from $N_u$ by substituting  $0 \mapsto 1$, $1 \mapsto e^\epsilon$ and taking normalization so that each row sums to one, as in Example \ref{ex:42ss}: 
    \begin{multline}
        Q_1 = \frac{1}{e^\epsilon+1}\begin{pmatrix}
        e^\epsilon & 1 \\
        e^\epsilon & 1 \\
        1 & e^\epsilon \\
        1 & e^\epsilon 
        \end{pmatrix},\;
        Q_2 =  \frac{1}{e^\epsilon+1}\begin{pmatrix}
        e^\epsilon & 1 \\
        1 & e^\epsilon \\
        e^\epsilon & 1 \\
        1 & e^\epsilon 
        \end{pmatrix},
        \\
        Q_3 = \frac{1}{e^\epsilon+1}\begin{pmatrix}
        e^\epsilon & 1 \\
        1 & e^\epsilon \\
        1 & e^\epsilon \\
        e^\epsilon & 1 
        \end{pmatrix}.
    \end{multline}
    Now, {for each $u \in \mathcal{U}$, let $\mathcal{Y}_u = \{u,6-u\}$, $f_u(y) = M_y$, $U$ be the uniform random variable over $\mathcal{U}$, and a conditional distribution $Q(\cdot|u,\cdot):\mathcal{X} \rightarrow \mathcal{Y}_u$ be $Q_u$.}
    Then, $(P_U,Q)$ becomes a decomposition of the $(4,2,\epsilon)$-SS mechanism with $C(P_U,Q) = 1$, which is less than {the communication cost  $\log 6$ of $(4,2)$-SS.}
\end{example}

By generalizing the above example, we propose a method of designing a communication-efficient $\epsilon$-LDP mechanism with shared randomness based on a resolution of a block design.

\begin{theorem}\label{thm:BlockDecomp}
Let $\tilde{Q}$ be a block design mechanism whose underlying block design has a resolution.
Then, there exists a decomposition $(P_U,Q)$ of $\tilde{Q}$ such that
\begin{equation}
    C(P_U,Q) = C(\tilde{Q}) - H(P_U).
\end{equation}
\end{theorem}

\begin{IEEEproof}
    Let $\tilde{Q}$ be an $\epsilon$-LDP block design mechanism based on a block design $(\mathcal{X},\mathcal{Z},\mathcal{I})$, and $\mathcal{R}=\{\mathcal{C}_1,\cdots,\mathcal{C}_{|\mathcal{R}|}\}$ be a resolution of $(\mathcal{X},\mathcal{Z},\mathcal{I})$.
    First, we construct $(P_U,Q)$ as follows:
    Let $\mathcal{U}=[1:|\mathcal{R}|]$ and  $P_U(u) = |\mathcal{C}_u|/|\mathcal{Z}|$ for each $u \in \mathcal{U}$.
    Let $M$ be an incidence matrix of $(\mathcal{X},\mathcal{Z},\mathcal{I})$ and $\mathcal{Y} = {[1:|\mathcal{Z}|]}$.
    Clearly, there exists a bijection $\psi:\mathcal{Y} \rightarrow \mathcal{Z}$, and let $\mathcal{Y}_u =\psi^{-1}(\mathcal{C}_u)$.
    Then, define $f_u:\mathcal{Y}_u \rightarrow \mathcal{Z}$ as a restriction of $\psi$ to $\mathcal{Y}_u$.
    Now, we define $Q(y|u,x)$ as
    \begin{multline}
    Q(y|u,x) 
    \\ = \begin{cases}
            \frac{e^\epsilon}{\alpha(\mathcal{C}_u)(e^\epsilon-1) + |\mathcal{C}_u|} & \text{ if } \psi(y) \in \mathcal{I}_x, \psi(y) \in \mathcal{C}_u
    \\  \frac{1}{\alpha(\mathcal{C}_u)(e^\epsilon-1) + |\mathcal{C}_u|} &           \text{ if }\psi(y) \notin \mathcal{I}_x, \psi(y) \in \mathcal{C}_u\\
            0 & \text{ otherwise}
        \end{cases}\label{eq:res_Q},
    \end{multline}
    for all $u\in \mathcal{U}, x\in \mathcal{X}, y\in \mathcal{Y}$.

    To check whether $(P_U,Q)$ is a decomposition of $\tilde{Q}$ or not, it is sufficient to check \eqref{eq:decomposition} because $(P_U,Q)$ clearly satisfies the $\epsilon$-LDP constraint, and $f_u$ is an injection for all $u \in \mathcal{U}$.
    Suppose $(\mathcal{X},\mathcal{Z},\mathcal{I})$ is a $(v,b,r,k,\lambda)$-block design.
    Because $\mathcal{R}$ is a partition of $\mathcal{Z}$ and $\mathcal{C}_u$ is the image of the injection $f_u$ for each $u \in \mathcal{U}$, we can  conclude that for each $z \in \mathcal{Z}$, there exists a unique pair $(u_z,y_z)$ such that $f_{u_z}(y_z) = z$.
    Thus,
    \begin{align}
        \sum_{u \in \mathcal{U}} \sum\limits_{\substack{y \in \mathcal{Y}_u : \\ f_u(y) = z}}& Q(y|u,x)  P_U(u) = Q(y_z|u_z,x) P_U(u_z)
        \\&=  \frac{(e^\epsilon-1) \mathbbm{1}((x,z) \in \mathcal{I}) + 1}{(e^\epsilon-1)b\alpha(\mathcal{C}_{u_z})/|\mathcal{C}_{u_z}| + b} \label{eq:thm_pf_1}
        \\& = \frac{(e^\epsilon-1) \mathbbm{1}((x,z) \in \mathcal{I}) + 1}{(e^\epsilon-1)bk/v + b} \label{eq:pf_step1}
        \\& = \frac{(e^\epsilon-1) \mathbbm{1}((x,z) \in \mathcal{I}) + 1}{(e^\epsilon-1)r + b} \label{eq:pf_step2}
        \\& = \tilde{Q}(z|x),
    \end{align}
    where \eqref{eq:thm_pf_1} follows from the construction of $(P_U,Q)$, \eqref{eq:pf_step1} from Lemma \ref{lem:res_lem}, \eqref{eq:pf_step2} from Lemma \ref{lem:blockDesignFundEq}, and the last equation from Definition \ref{def:blockDesignMech}.

    The communication cost can be calculated as 
    \begin{align}
        C(P_U & ,Q) = \sum_{u \in \mathcal{U}} P_U(u) \log |\mathcal{Y}_u| = \sum_{u \in \mathcal{U}} P_U(u) \log |\mathcal{C}_u| \label{eq:comm_res_simple}
        \\&= \sum_{u \in \mathcal{U}} P_U(u) \log (b P_U(u))
        \\&= (\log b) \sum_{u \in \mathcal{U}} P_U(u) + \sum_{u \in \mathcal{U}} P_U(u) \log P_U(u)
        \\&= \log b - H(P_U),
    \end{align}
    because $\mathcal{C}_u$ is the image of the injection $f_u:\mathcal{Y}_u \rightarrow \mathcal{Z}$.
\end{IEEEproof}

Note that the construction of $(P_U,Q)$ and its communication cost in the proof depend on the choice of  resolution $\mathcal{R}$.
We call a decomposition $(P_U,Q)$ of $\tilde{Q}$ constructed in the proof of Theorem \ref{thm:BlockDecomp} as a \textbf{resolution of a block design mechanism} by $\mathcal{R}$.
For any given block design scheme $(\tilde{Q},\tilde{P}_n)$ and a resolution $(P_U,Q)$ of $\tilde{Q}$, an estimator $\hat{P}_n$ for $(P_U,Q)$ can be constructed in a straightforward manner so that $L_n(v,(P_U,Q),\hat{P}_n) = L_n(v,\tilde{Q},\tilde{P}_n)$ as in Lemma \ref{lem:est_dec}.
We call such $((P_U,Q),\hat{P})$ as a \textbf{resolution of a block design scheme} $(\tilde{Q},\tilde{P})$.

For the case when a resolution of a block design scheme is constructed by an $\alpha$-resolution, its communication cost can be calculated as follows.
\begin{corollary}\label{cor:comm_alpha_res}
    Let $(P_U,Q)$ be a resolution of a block design scheme $\tilde{Q}$ with parameters $(v,k)$ by an $\alpha$-resolution $\mathcal{R}$.
    Then, $C(P_U,Q) = v \alpha / k$.
\end{corollary}
\begin{IEEEproof}
    From Lemma \ref{lem:res_lem}, we have $|\mathcal{C}| = v \alpha / k$ for any $\mathcal{C} \in \mathcal{R}$.
    Combining with \eqref{eq:comm_res_simple}, we get the desired result.
\end{IEEEproof}

\section{Resolutions of Subset-Selection Scheme} \label{sec:res_SS}
Since the SS with the  parameters $(v,k^*,\epsilon)$ is a block design scheme which achieves the exactly optimal PUT \cite{ye18_SS,ye19_SSOptim, park23_block}, its resolution also achieves the exactly optimal PUT while requiring less communication cost in general.
In this section, we provide two resolutions of the optimal SS constructed from Baranyai's theorem \cite{Baranyai74_Baranyai} and from the cyclic shift group action.

\subsection{Resolution from Baranyai's theorem}\label{subsec:Baranyai}
The following theorem states that the minimum communication cost among all possible choices of PUT-optimal block design and its resolution can be achieved by a resolution of the optimal SS.
\begin{theorem} \label{thm:BarSS}
    For any $(v,\epsilon)$, there exists a resolution of the optimal SS with parameters $(v,k^*,\epsilon)$, whose communication cost is $\log \frac{v}{\mathrm{gcd}(v,k^*)}$-bits.
    Moreover, the communication cost of a resolution of a block design scheme achieving the optimal PUT cannot be less than $\log \frac{v}{\mathrm{gcd}(v,k^*)}$-bits.
\end{theorem}
{\begin{remark}\label{rmk:min_cost}
    By Proposition \ref{prop:optimal_eqcond}, $k^* = v/2$
    when $v$ is even and $\epsilon \leq \frac{1}{2} \ln \frac{v + 2}{v - 2}$.
    In this case, Theorem \ref{thm:BarSS} implies that the optimal PUT can be achieved by a scheme with the communication cost of 1-bit.
    In other words, for any even $v$, the communication cost of 1-bit is sufficient to achieve the optimal PUT for a high privacy regime.
\end{remark}}
The proof of Theorem \ref{thm:BarSS} is based on Baranyai's theorem \cite{Baranyai74_Baranyai}, which is about a factorization of a complete uniform hypergraph.
There is a natural one-to-one correspondence between incidence structures and hypergraphs. According to this correspondence, a complete uniform hypergraph corresponds to a complete block design, and a factorization of a hypergraph corresponds to a resolution of an incidence structure.
In the following, we state Baranyai's theorem in terms of incidence structure and present its corollaries.

\begin{theorem}[Baranyai's theorem]\label{thm:Baranyai}
    Let $(\mathcal{X},\mathcal{Z},\mathcal{I})$ be a $(v,k)$-complete block design.
    If $c_1,\cdots,c_t$ are natural numbers satisfying $\sum\limits_{i=1}^t c_i = \binom{v}{k}$, then there exists a partition $\mathcal{R}=\{\mathcal{C}_1,\cdots,\mathcal{C}_t\}$ of $\mathcal{Z}$ such that for all $i \in [1:t]$, $|\mathcal{C}_i|=c_i$ and 
    \begin{equation}
        \forall x \in \mathcal{X},\quad |C_i \cap \mathcal{I}_x| \in \left\{ \left\lfloor \frac{c_i k}{v} \right\rfloor, \left\lceil \frac{c_i k}{v} \right\rceil \right\}.
    \end{equation}
\end{theorem}

\begin{corollary}\label{cor:min_fac}
    There exists a $\frac{k}{\mathrm{gcd}(v,k)}$-resolution of a $(v,k)$-complete block design.
    Moreover, for any resolution class $\mathcal{C}$ of a block design with parameters $(v,k)$, $\alpha(\mathcal{C}) \geq \frac{k}{\mathrm{gcd}(v,k)}$.
\end{corollary}
\begin{IEEEproof}
    Substituting $t=\frac{\mathrm{gcd}(v,k)}{v}\binom{v}{k}$ and $c_i = \binom{v}{k}/t$ for all $i \in [1:t]$ in Theorem \ref{thm:Baranyai} implies the existence of a $\frac{k}{\mathrm{gcd}(v,k)}$-resolution. Next, Lemma \ref{lem:res_lem} implies that $\alpha(\mathcal{C})$ must be a multiple of $\frac{k}{\mathrm{gcd}(v,k)}$.    
\end{IEEEproof}


Now, let us prove Theorem \ref{thm:BarSS} based on Corollary \ref{cor:min_fac}.
\begin{IEEEproof}[Proof of Theorem \ref{thm:BarSS}]
    For given $(v,\epsilon)$, the optimal SS is the $(v,k^*,\epsilon)$-SS.
    By Corollary \ref{cor:min_fac}, there exists a $\frac{k^*}{\mathrm{gcd}(v,k^*)}$-resolution $\mathcal{R}$ of the $(v,k^*)$-complete block design.
    Thus, there exists $(P_U,Q)$, the resolution of the optimal SS by $\mathcal{R}$.
    The communication cost directly follows from Corollary \ref{cor:comm_alpha_res}.
    
    For the last claim, let $(P_U,Q)$ be a resolution of a block design scheme by $\mathcal{R}$ achieving the optimal PUT, and suppose that the block design scheme has parameters $(v,k)$.
    By Proposition \ref{prop:optimal_eqcond}, $k \in K_{opt}(v,\epsilon)$.
    Also, for any $\mathcal{C} \in \mathcal{R}$, $\alpha(\mathcal{C}) \geq \frac{k}{\mathrm{gcd}(v,k)}$ by Corollary \ref{cor:min_fac}, and Lemma \ref{lem:res_lem} implies $|\mathcal{C}| \geq \frac{v}{\mathrm{gcd}(v,k)}$.
    Therefore, 
    \begin{align}
        C(P_U,Q) &= \sum\limits_{u \in \mathcal{U}} P_U(u) \log |\mathcal{C}_u|\\
        &\geq \min\limits_{k \in K_{opt}(v,\epsilon)} \log \frac{v}{\mathrm{gcd}(v,k)} 
        \\ &= \log \frac{v}{\mathrm{gcd}(v,k^*)}.
    \end{align}
\end{IEEEproof}
We call a resolution of SS constructed in the same way as in the proof of Theorem \ref{thm:BarSS} a \textbf{Baranyai's resolution} of SS.
Note that a Baranyai's resolution of the optimal $(v,k^*,\epsilon)$-SS achieves the optimal PUT and its communication cost is the minimum among the communication costs of resolutions of block design mechanisms achieving the optimal PUT. 
However, a Baranyai's resolution of SS can be obtained from a recursive algorithm in general because known proofs of Baranyai's theorem are based on the maximum flow over a flow network \cite{Baranyai74_Baranyai,chee21_Baranyai_const4}. 
In the following subsection, we suggest another resolution of SS based on the cyclic shift group action which has an explicit structure.

\subsection{Resolution from the cyclic shift group action}\label{subsec:cyclic}
As stated in Remark \ref{rmk:CBD}, in our representation of $(v,k)$-complete block design $(\mathcal{X},\mathcal{Z},\mathcal{I})$, $\mathcal{Z}$  is the set all $\{0,1\}$-valued vectors of length $v$ whose Hamming weight is $k$.
Thus, for every $z \in \mathcal{Z}$, any cyclic shifted version of $z$ should be also contained in $\mathcal{Z}$.
Therefore, the cyclic shift operations on $\mathcal{Z}$ gives a group action.
From this fact, the orbits of the cyclic shift group action on $\mathcal{Z}$ form a partition of $\mathcal{Z}$.
Further, we show that such a set of orbits is also a resolution of a complete block design, which results in a resolution of SS since SS is based on a complete block design.
We first provide a simple example which shows how a resolution of SS can be constructed from the cyclic shift group action.
\begin{example}
    Let $M$ be an incidence matrix of a $(4,2)$-complete block design $(\mathcal{X},\mathcal{Z},\mathcal{I})$ as in Example \ref{ex:42ss}, and note that the set of columns of $M$ is equal to $\mathcal{Z}$.
    Then, the orbits of the cyclic shift group action on $\mathcal{Z}$ are
    $\mathcal{Z}_1 = \{ M^1, M^4, M^6, M^3 \}$ and $\mathcal{Z}_2 = \{M^2, M^5\}$, where $M^y$ denotes the $y$-th column of $M$.
    It can be easily checked that $\{ \mathcal{Z}_1, \mathcal{Z}_2 \}$ is a resolution of $(\mathcal{X},\mathcal{Z},\mathcal{I})$.
    As in Example \ref{ex:res_block}, we can get row stochastic matrices $Q_1$ and $Q_2$ from $\mathcal{Z}_1$ and $\mathcal{Z}_2$, respectively, i.e.,
    \begin{align}
        Q_1 &= \frac{1}{2(e^\epsilon+1)}\begin{pmatrix}
        e^\epsilon & 1 & 1 & e^\epsilon \\
        e^\epsilon & e^\epsilon & 1 & 1 \\
        1 & e^\epsilon & e^\epsilon & 1 \\
        1 & 1 & e^\epsilon & e^\epsilon 
        \end{pmatrix},
        \\ Q_2 &= \frac{1}{e^\epsilon+1}\begin{pmatrix}
        e^\epsilon & 1 \\
        1 & e^\epsilon \\
        e^\epsilon & 1 \\
        1 & e^\epsilon 
        \end{pmatrix}.
    \end{align}
    If we let $\mathcal{U}=[1:2]$, $\mathcal{Y}_1 = \{1,4,6,3\}$, $\mathcal{Y}_2 = \{2,5\}$, $f_u(y) = M_y$, $U \sim P_U$ with $P_U(1) = 2/3$ and  $P_U(2) = 1/3$, and $Q(\cdot|u,\cdot):\mathcal{X} \rightarrow \mathcal{Y}_u$ be $Q_u$, then $(P_U,Q)$ is a resolution of the $(4,2,\epsilon)$-SS with $C(P_U,Q) = 5/3$-bits.
\end{example}

To give a formal description of a resolution of SS by the cyclic shift group action, it is convenient to start with a general group action.
Let $(\mathcal{X},\mathcal{Z},\mathcal{I})$ be a complete block design, $G$ be a finite group, and suppose that $G$ acts on $\mathcal{X}$.
Then, there is an induced group action of $G$ on $\mathcal{Z}$,
\begin{equation}\label{eq:induced_g_act}
    \forall g \in G, \; x \in \mathcal{X},\; z \in \mathcal{Z}, \quad (g(z))_x := z_{g^{-1}(x)}.
\end{equation}
Thus, $\mathcal{Z}$ can be partitioned into the orbits of the induced group action.
We show that such a partition is a resolution of a complete block design when $G$ acts on $\mathcal{Z}$ transitively.
\begin{proposition}\label{prop:cyc_res}
    Let $(\mathcal{X},\mathcal{Z},\mathcal{I})$ be a complete block design, $G$ be a finite group, and suppose $G$ acts on $\mathcal{X}$.
    If a group action of $G$ on $\mathcal{X}$ is transitive, then the set of all orbits $\mathcal{R} = \mathcal{Z} / G$ of the induced group action in \eqref{eq:induced_g_act} is a resolution of $(\mathcal{X},\mathcal{Z},\mathcal{I})$. 
\end{proposition}
\begin{IEEEproof}
    Let $\mathcal{C} \in \mathcal{Z} / G$, $\bar{z} \in \mathcal{C}$, and $G_{\bar{z}}$ be the stabilizer of $\bar{z}$.
    By the orbit-stabilizer theorem \cite[Prop. 2, Chap. 4]{dummit04_algebra}, there is a one-to-one correspondence between $\mathcal{C}$ and the set of all cosets of $G_{\bar{z}}$ in $G$, by identifying $g(\bar{z}) \leftrightarrow g G_{\bar{z}}$.
    As a result, for $x \in \mathcal{X}$, we have
    \begin{align}
        |\mathcal{C} \cap \mathcal{I}_x| &= \sum_{z \in \mathcal{C}}z_x = \frac{1}{|G_{\bar{z}}|} \sum_{g \in G}(g(\bar{z}))_x \label{eq:circ_pf_1}
        \\&= \frac{1}{|G_{\bar{z}}|} \sum_{g \in G} \bar{z}_{g^{-1}(x)},
    \end{align}
    where the last equation follows from \eqref{eq:induced_g_act}.
    By the transitivity, for any $x' \in \mathcal{X}$, there exists $g' \in G$ such that $x' = g'(x)$.
    Thus, we have
    \begin{align}
        \frac{1}{|G_{\bar{z}}|} \sum_{g \in G} \bar{z}_{g^{-1}(x)} &= \frac{1}{|G_{\bar{z}}|} \sum_{g \in G} \bar{z}_{(g' g)^{-1}(x')}
        \\& = \frac{1}{|G_{\bar{z}}|} \sum_{g \in G} \bar{z}_{g^{-1}(x')} = |\mathcal{C} \cap \mathcal{I}_{x'}|.\label{eq:circ_pf_2}
    \end{align}
    Through \eqref{eq:circ_pf_1}-\eqref{eq:circ_pf_2}, we have $|\mathcal{C} \cap \mathcal{I}_x| = |\mathcal{C} \cap \mathcal{I}_{x'}|$ for all $x,x' \in \mathcal{X}$, which implies that $\mathcal{C}$ is a resolution class of $\mathcal{Z}$.
    Therefore, $\mathcal{R} = \mathcal{Z}/G$ is a resolution of $(\mathcal{X},\mathcal{Z},\mathcal{I})$.
\end{IEEEproof}

Let $(\mathcal{X},\mathcal{Z},\mathcal{I})$ be a complete block design.
The simplest transitive group action on $\mathcal{X}$ is the modulo sum action.
Let $G=\mathbb{Z}/v\mathbb{Z}$ and define a group action of $G$ on $\mathcal{X}$ by $g(x) = (g+(x-1) \mod v) + 1$.
Then, the induced group action of $G$ on $\mathcal{Z}$ is just the cyclic shift group action on the set of vectors, $\mathcal{Z}$.
By Proposition \ref{prop:cyc_res}, the cyclic shift group action constructs a resolution of a complete block design, and it induces a resolution of SS.
We call such resolution as a \textbf{cyclic shift resolution} of SS.
The communication cost of a cyclic shift resolution of the $(v,k,\epsilon)$-SS is at most $\log v$-bits, because the size of each orbit is at most $|G|=v$.
The exact communication cost can be calculated as follows, whose proof is in the appendix.
\begin{proposition}\label{prop:cyclicShiftComCost}
The communication cost of a cyclic shift resolution of $(v,k)$-SS is
\begin{equation}
    \log v - \frac{1}{\binom{v}{k}} \sum_{\substack{p|\gcd(v,k)\\p \text{\normalfont{ is prime}}}} \sum_{i=1}^{\beta_p} \binom{v/p^i}{k/p^i} \log p,
\end{equation}
where $\beta_p$ is the largest integer $\beta$ such that $p^\beta | \gcd(v, k)$.
\end{proposition}

\section{Other Resolutions}\label{sec:other}
The Baranyai's resolution of an optimal SS in Section~\ref{subsec:Baranyai} always exists and has the minimum communication cost among all the possible resolutions of PUT-optimal block design schemes, but it is obtained from a recursive algorithm in general.
On the other hand, the cyclic shift resolution of an optimal SS in Section \ref{subsec:cyclic} always exists with an explicit structure, but its communication cost can be larger than the Baranyai's resolution.

In this section, we provide resolutions of some other block design schemes than SS, which have explicit structures and achieve the optimal PUT with the communication cost equal to the Baranyai's resolution of the optimal SS for some values of $(v,\epsilon)$.
The main idea is to exploit an $\alpha$-resolvable block design because a resolution of a block design scheme with parameters $(v,k)$ by $\alpha$-resolution has the communication cost of $\log \frac{v\alpha}{k}$-bits by Corollary \ref{cor:comm_alpha_res}, {and it has the communication cost equal to the Baranyai's resolution when $\alpha = k/\mathrm{gcd}(v,k)$.
Note that if a block design with parameters $(v,k)$ is resolvable, then $k|v$ by Lemma \ref{lem:res_lem}.}
In this paper, we exploit two resolvable block designs, affine geometry \cite[Sec. 5.2.3]{stinson04_combinatorial} and Hadamard 3-design \cite{Hedayat78_Hadamard}, as representative examples.

\subsection{Resolution from affine geometry}
An affine geometry is a resolvable block design constructed by affine subspaces in a finite field $\mathbb{F}_q^d$ for a prime power $q$ and a positive integer $d$ \cite[Sec. 5.2.3]{stinson04_combinatorial}.
\begin{definition}
    For a prime power $q$ and integers ${d > m \geq 1}$, an affine geometry $\mathrm{AG}_m(d,q)$ is an incidence structure $(\mathcal{X},\mathcal{Z},\mathcal{I})$ such that
    \begin{itemize}
        \item $\mathcal{X} = \mathbb{F}_q^d$.
        \item $\mathcal{Z}$ is the set of all affine subspaces of dimension $m$ in $\mathcal{X}$, i.e., cosets of $m$-dimensional linear subspaces in $\mathcal{X}$.
        \item $\mathcal{I}=\{(x,z) \in \mathcal{X}\times \mathcal{Z}:x \in z\}$.
    \end{itemize}
\end{definition}

\begin{lemma}\cite[Thm. 5.12]{stinson04_combinatorial}\label{lem:affine}
    $\mathrm{AG}_m(d,q)$ is a block design with parameters
    \begin{multline}
        (v,b,r,k,\lambda)
        \\ = \left( q^d, q^{d-m} \binom{d}{m}_q, \binom{d}{m}_q, q^m, \binom{d-1}{m-1}_q \right),
    \end{multline}
    where
    \begin{equation}
        \binom{d}{m}_q := \frac{(q^d-1)(q^{d-1}-1)\cdots(q^{d-m+1}-1)}{(q^m-1)(q^{m-1}-1)\cdots(q-1)}.
    \end{equation}
    Moreover, $\mathrm{AG}_m(d,q)$ has a 1-resolution which is the set of all parallel classes, where a parallel class is a set of all parallel affine subspaces.\footnote{Two affine subspaces $A$, $B$ are said to be parallel if one of them is a coset of the other one.}
\end{lemma}

Because $\mathrm{AG}_m(d,q)$ is a resolvable block design, we can get the resolution of the block design constructed from $\mathrm{AG}_m(d,q)$ by the 1-resolution of $\mathrm{AG}_m(d,q)$.
Its communication cost and a necessary and sufficient condition for achieving the optimal PUT are as follows.
\begin{corollary}
    Let $(\tilde{Q},\tilde{P}_n)$ be a block design scheme whose underlying block design is $\mathrm{AG}_m(d,q)$, and $((P_U,Q),\hat{P}_n)$ be the resolution of $(\tilde{Q},\tilde{P}_n)$ by the 1-resolution of $\mathrm{AG}_m(d,q)$.
    Then,
    \begin{equation}
        C(P_U,Q) = (d-m) \log q,
    \end{equation}
    and $((P_U,Q),\hat{P}_n)$ achieves the optimal PUT if and only if
    \begin{equation}\label{eq:opt_eps_AG}
        \epsilon \in [E(q^m,q^m+1;q^d), E(q^m-1,q^m;q^d)].
    \end{equation}
\end{corollary}
\begin{IEEEproof}
    The communication cost is equal to $\log \frac{v}{k}$-bits as we mentioned in the beginning of this section, and Lemma \ref{lem:affine} gives the desired result.
    The optimality condition \eqref{eq:opt_eps_AG} directly follows from Proposition \ref{prop:optimal_eqcond}.
\end{IEEEproof}
Note that the optimality condition \eqref{eq:opt_eps_AG} reduces to $\epsilon \approx {(d-m) \ln q} = \ln \frac{v}{k}$ when $q \gg 1$.

\subsection{Resolution from Hadamard 3-design}
A Hadamard 3-design is a resolvable block design constructed through some manipulations on a normalized Hadamard matrix.
The details about a Hadamard matrix and a Hadamard 3-design can be found in \cite{ionin06_combSymDesign,Hedayat78_Hadamard}.
\begin{definition}
    A Hadamard matrix of order $m$ is an $m \times m$ matrix whose entries are in $\{-1,1\}$ and whose rows are mutually orthogonal.
    A Hadamard matrix is called normalized if every entry of the first column and the first row is 1.
    For a normalized Hadamard matrix $H$, the core of $H$ is the submatrix of $H$ obtained by removing its first column and first row.
\end{definition}
Note that the existence of a Hadamard matrix of order $m$ is equivalent to the existence of a normalized Hadamard matrix of order $m$, and there is the well-known conjecture called Hadamard conjecture which claims that a Hadamard matrix of order $4t$ exists for all positive integer $t$.
For a given normalized Hadamard matrix, a Hadamard 3-design can be constructed as follows.
\begin{lemma}\cite[Thm. 4.2.]{Hedayat78_Hadamard}
    Let $t$ be a positive integer.
    If there exists a normalized Hadamard matrix $H$ of order $4t$, then there exists a resolvable block design $H_3(t)$ with parameters $(4t, {8t-2}, 4t-1, 2t, 2t-1)$, which is called a Hadamard 3-design.
    An incidence matrix of $H_3(t)$ is given by
    \begin{equation}\label{eq:inc_H3}
        \begin{pmatrix} H_1 & H_2 \\ \mathbf{1} & \mathbf{0}\end{pmatrix} \in \{0,1\}^{4t \times (8t-2)},
    \end{equation}
    where $H_1=\frac{1}{2}(J+A)$, $H_2=\frac{1}{2}(J-A)$, $A$ is the core of $H$, and $J$ is the $(4t-1) \times (4t-1)$ matrix whose entries are all one.
\end{lemma}
Note that a normalized Hadamard matrix of order $4t$ has an explicit construction for some cases: when $t$ is a power of 2, there is Sylvester's construction, and when $(4t-1)$ is a prime power, there is Paley's construction.
The 1-resolution of $H_3(t)$ can be simply represented in terms of its incidence matrix as follows.
Let $M$ be the matrix in \eqref{eq:inc_H3} and $M^y$ be the $y$-th column of $M$. 
Then, $\{\{M^i,M^{i+4t-1}\}:i \in [1:4t-1]\}$ corresponds to the 1-resolution of $H_3(t)$.
The resolution of the block design mechanism constructed from $H_3(t)$ by the 1-resolution has the communication cost of only 1-bit, and it achieves the optimal PUT for a high privacy regime.

\begin{corollary}\label{cor:H3_res}
    Let $(\tilde{Q},\tilde{P}_n)$ be a block design scheme whose underlying block design is $H_3(t)$, and $((P_U,Q),\hat{P}_n)$ be the resolution of $(\tilde{Q},\tilde{P}_n)$ by the 1-resolution of $H_3(t)$.
    Then, $C(P_U,Q) = 1$ and $((P_U,Q),\hat{P}_n)$ achieves the optimal PUT if and only if $\epsilon \leq \frac{1}{2}\ln\frac{2t+1}{2t-1}$.
\end{corollary}
\begin{IEEEproof}
    Because $H_3(t)$ is a block design with parameters $(v,k) = (4t,2t)$, $C(P_U,Q) = \log v/k = 1$.
    Also, by Proposition \ref{prop:optimal_eqcond}, the optimality condition is
    \begin{equation}
        \epsilon \in [E(2t,2t+1;4t),E(2t-1,2t;4t)].
    \end{equation}
    Through simple calculations, it can be checked that $E(2t,{2t+1};4t) \leq 0$ and
    $E(2t-1,2t;4t) = \frac{1}{2} \ln \frac{2t+1}{2t-1}$.
\end{IEEEproof}

\begin{remark}
    Not only the block designs and their resolutions provided in this paper, lots of block designs and their resolutions are known \cite{ionin06_combSymDesign,colbourne07_handbook}, and the resultant resolutions of block design schemes can be considered.
    For example, for any prime power $q$, there exists a resolvable $({q^3+1}, {q^2(q^3+1)/(q+1)}, q^2, q+1, 1)$-block design \cite[Thm. 5.3.9]{ionin06_combSymDesign}.
    For $v= q^3+1 \gg 1$, the resolution of the block design scheme constructed from this resolvable block design achieves the optimal PUT when $\epsilon \approx \frac{1}{3}\ln v$ with the communication cost of $\frac{1}{3}\log v$, approximately.
    Also, for any even $v$ and any $\lambda \geq 1$, there exists a resolvable $(v, \lambda v(v-1)/2 , \lambda(v-1), 2, \lambda)$-block design called a round robin tournament \cite[Prop. 5.3.5]{ionin06_combSymDesign}.
    For $v \gg 1$, the resolution of the block design scheme constructed from the round robin tournament achieves the optimal PUT when $\epsilon \approx \ln v$ with the communication cost of $\log v$, approximately.
\end{remark}

\section{Conclusion}\label{sec:conc}
In this paper, we suggested a novel way of exploiting shared randomness to achieve the exactly optimal PUT of discrete distribution estimation problem with lower communication costs.
The key idea was to decompose a block design scheme based on a resolution of the underlying block design.
The resultant scheme decomposed in this way achieves the same PUT as the original block scheme, while requiring less communication cost.
We proposed two decompositions of the exactly PUT-optimal SS, based on the Baranayi's theorem and the cyclic shift group action, both requiring the communication costs of at most the input data size. The former achieves the minimum communication cost among the resolution-based decompositions of PUT-optimal block design schemes, and it can be constructed through a recursive algorithm. The latter has an explicit structure, but it requires more communication cost than the former in general. One can consider decomposing other block design schemes, as we exemplified with affine geometry and Hadamard 3-design.

An interesting future work would be characterizing the exactly optimal privacy-utility-communication trade-off with shared randomness, not in the order-optimal sense, which subsumes the problem of finding the minimum communication cost to achieve the optimal PUT.
 We partially answered this problem, i.e., according to Remark \ref{rmk:shared_rand}, when $v$ is even and $\epsilon$ is small enough, the communication cost of 1-bit is sufficient to achieve the optimal PUT, and it is in fact the minimum communication cost.
 However, when $v$ is odd, our resolution-based decompositions of block design schemes require the communication costs of more than 1-bit to achieve the optimal PUT for all privacy regime.
 Intuitively, we expect that for a high privacy regime, the communication cost of 1-bit would be sufficient to achieve the optimal PUT, which may be possible by taking other approaches of exploiting shared randomness.



\appendix
\begin{IEEEproof}[Proof of Proposition \ref{prop:cyclicShiftComCost}]
    For a notational simplicity, we treat a number in $[0:v-1]$ as both an integer and an element of $\mathbb{Z}/v\mathbb{Z}$.
    First, we classify the elements of $\mathcal{Z}$ by their stabilizer.
    Note that the complete set of all subgroups of $G=\mathbb{Z}/v\mathbb{Z}$ is $\{\langle d \rangle : d | v\}$, where we identify $\langle v \rangle = \langle 0 \rangle$.
    For each $d$ satisfying $d|v$, let's define the following numbers:
    \begin{itemize}
        \item $s_d$ is the number of $z \in \mathcal{Z}$ which is stabilized by $\langle d \rangle$, or equivalently, the number of $z \in \mathcal{Z}$ such that the $d$-times cyclic shifted version of $z$ is also $z$.
        \item $n_d$ is the number of $z \in \mathcal{Z}$ whose stabilizer is precisely $\langle d \rangle$, or equivalently, the number of $z \in \mathcal{Z}$ such that $d$ is the smallest positive integer satisfying that the $d$-times cyclic shifted version of $z$ is also $z$.
    \end{itemize}
    Note that $s_d = \sum_{e:e | d} n_e$. 
    
    Next, we calculate $s_d$.
    By definition, $z \in \mathcal{Z}$ is stabilized by $\langle d \rangle$ if and only if for each $i=0,1,\cdots,(v/d)-1$, $z_{x} = z_{[(x-1) + id \mod v] + 1}$ holds for all $x=1,2,\cdots,v$.
    In other words,
    \begin{align}
        z_{j} = z_{d+j} = z_{2d+j} = \cdots = z_{(v/d-1)d+j}, \label{eq:stabCond}
    \end{align}
    for all $j=1,2,\cdots,d$.
    As a result, if $z \in \mathcal{Z}$ is stabilized by $\langle d \rangle$, then the number of $1$'s in $z$, which is $k$, must be a multiple of $v/d$. Hence, if $(v/d)$ does not divide $k$, then $s_d = 0$.
    Also, if $(v/d)$ divides $k$, then an element $z \in \mathcal{Z}$ stabilized by $\langle d \rangle$ is completely characterized by $(z_1,z_2,\cdots,z_d)$, which have $k/(v/d)=kd/v$ number of ones.
    Thus, $s_d$ is equal to the number of ways to choose $kd/v$ indices in $\{1,2,\cdots,d\}$, i.e.,
    \begin{equation}
            s_d = \binom{d}{kd/v}.
    \end{equation}
    By the simple reparameterization, we get
    \begin{equation}\label{eq:s_reparam}
        s_{v/d} = \begin{cases}
            \binom{v/d}{k/d} & \text{if }d | k\\
            0 & \text{otherwise}\\
        \end{cases}.
    \end{equation}

    Now, we calculate the communication cost,
    \begin{align}
        \sum_{u \in \mathcal{U}} P_U(u) \log |\mathcal{Y}_u| &= \sum_{u \in \mathcal{U}} \frac{|\mathcal{C}_u|}{b} \log |\mathcal{C}_u|\\
        &= \sum_{u \in \mathcal{U}} \sum_{z \in \mathcal{C}_u} \frac{1}{b} \log |\mathcal{C}_u|. \label{eq:comCost-Prev_orbitStab}
        \end{align}
        By the orbit-stabilizer theorem \cite[Prop. 2, Chap. 4]{dummit04_algebra}, we have $|\mathcal{C}_u|=|G|/|G_z|$ whenever $z \in \mathcal{C}_u$.
        Hence, \eqref{eq:comCost-Prev_orbitStab} becomes
        \begin{align}
        \sum_{u \in \mathcal{U}} \sum_{z \in \mathcal{C}_u} \frac{1}{b} \log \frac{|G|}{|G_z|} &= \sum_{z \in \mathcal{Z}} \frac{1}{b}\log \frac{|G|}{|G_z|}\\
        &= \sum_{z \in \mathcal{Z}} \frac{1}{b}(\log v - \log |G_z|)\\
        &= \log v - \frac{1}{b}\sum_{d:d|v} n_d \log (v/d). \label{eq:comCost-Prev_MobiusInv}
        \end{align}
        Since $s_d = \sum_{e:e | d} n_e$ for each $d | v$, the M\"obius inversion formula gives $n_d = \sum_{e:e|d} s_e \mu(d/e)$ for each $d | v$, where $\mu$ is the M\"obius function\cite{burton11_elem_num_theory}.
        Hence, \eqref{eq:comCost-Prev_MobiusInv} becomes
        \begin{align}
        \log v - \frac{1}{b}\sum_{d:d|v} \sum_{e:e|d} s_e \mu(d/e) \log (v/d). \label{eq:comCost-Prev_changeVar}
        \end{align}
        Let $d'=v/d$, and note that $e|d|v$ can be equivalently written as $e|v$ and $d' = \frac{v}{d} | \frac{v}{e}$.
        Thus, \eqref{eq:comCost-Prev_changeVar} can be written as
        \begin{align}
        \log v - \frac{1}{b}\sum_{e:e|v} \sum_{d':d'|(v/e)} s_e \mu\left(\frac{v}{e}\frac{1}{d'}\right) \log (d'). \label{eq:comCost-Prev_Mangoldt}
        \end{align}
        Finally, using the Mangoldt function\cite{burton11_elem_num_theory}
        \begin{align}
            \Lambda(t) = \begin{cases} 
            0 & (\text{if } t \text{ is not a prime power}) \\
            \log p & (\text{if } t \text{ is a power of a prime } p)
            \end{cases},
        \end{align}
        which satisfies $\Lambda(t) = \sum_{d'|t} \mu(t/d') \log (d')$ for all positive integer $t$, \eqref{eq:comCost-Prev_Mangoldt} becomes
        \begin{align}
        &\log v - \frac{1}{b}\sum_{e:e|v} s_e \Lambda(v/e) \\
        &= \log v - \frac{1}{b}\sum_{e:e|v} s_{v/e} \Lambda(e)\\
        &= \log v - \frac{1}{b}\sum_{e:e|v, e|k} \binom{v/e}{k/e} \Lambda(e) \label{eq:pf_appen_last}\\
        &= \log v - \frac{1}{b}\sum_{e:e|\gcd(v,k)} \binom{v/e}{k/e} \Lambda(e)\\
        &= \log v - \frac{1}{\binom{v}{k}} \sum_{\substack{p|\gcd(v,k)\\p \text{ prime}}} \sum_{i=1}^{\beta_p} \binom{v/p^i}{k/p^i} \log p,
    \end{align}
    where \eqref{eq:pf_appen_last} follows from \eqref{eq:s_reparam}.
\end{IEEEproof}

\bibliographystyle{IEEEtran}
\bibliography{IEEEabrv, ref}

\end{document}